\documentclass[twocolumn,jcp,superscriptaddress,showpacs,floats,floatfix,preprintnumbers,amsmath,amssymb]{revtex4}
\usepackage{graphicx}
\usepackage{tabularx}
\usepackage[usenames]{color}
\usepackage{float}
\usepackage{subfig}
\usepackage{dcolumn}
\usepackage{bm}
\begin{document}
\title {The effect of ionization on the global minima of small and medium sized silicon and magnesium clusters}
\author{Sandip De}
\affiliation{Department of Physics, Universit\"{a}t Basel, Klingelbergstr. 82, 4056 Basel, Switzerland}
\author{S. Alireza Ghasemi}
\affiliation{Department of Physics, Universit\"{a}t Basel, Klingelbergstr. 82, 4056 Basel, Switzerland}
\author{Alexander Willand}
\affiliation{Department of Physics, Universit\"{a}t Basel, Klingelbergstr. 82, 4056 Basel, Switzerland}
\author{Luigi Genovese}
\affiliation{European Synchrotron Radiation Facility, 6 rue Horowitz, BP 220, 38043 Grenoble France}
\author{Dilip Kanhere}
\affiliation{
Department of Physics and
Center for Modeling and Simulation,
University of Pune,
Ganeshkhind,
Pune 411 007,
India}
\author{Stefan Goedecker}
\affiliation{Department of Physics, Universit\"{a}t Basel, Klingelbergstr. 82, 4056 Basel, Switzerland}
\date{\today}
\begin{abstract}
 We reexamine the question of whether the geometrical ground state of neutral and ionized clusters are identical.
Using a well defined criterion for being `identical' together extensive sampling methods on a potential energy 
surface calculated by density functional theory, 
we show that the ground states are in general different. This behavior is 
to be expected whenever there are meta-stable configurations which are close in energy to the ground state, but it 
disagrees with previous studies.

\end{abstract}
\maketitle
\section{Introduction}
Since experimental mass selection methods require ionized systems, the majority of experimental information on clusters was obtained for ionized clusters. On the other hand, neutral systems are of greater practical interest and the majority of theoretical works are done on neutral systems. The relation between the properties of neutral and 
ionized clusters is therefore an important one. The basic property which determines all other properties is the structure. 
Finding the global minimum structure of a cluster is a complex global geometry optimization problem on a high dimensional potential energy 
landscape~\cite{wales} with a huge number of local minima. In order to make accurate structural predictions, the potential energy surface should be 
calculated within density functional theory. Doing exhaustive unbiased searches for the global minimum at the density functional level 
has only recently become possible through the combined improvements in global optimization algorithms and computer performance. 

One basic question concerning the relation between neutral and ionized clusters is whether they have the same basic structure. Evidently adding or 
removing one electron will change the the exact bond lengths and angles but one might suspect that the structures remains nevertheless very similar. 
The relation between the structure of neutral and ionized clusters has been investigated in numerous previous publications 
for the same silicon and magnesium clusters that we have reexamined. 
The conclusion, in all the publications we are aware of, is that in general the structures of the neutral and cation clusters 
are more or less identical, but the criteria for being `identical' are not always explicitly given. 
We introduce a well defined criterion for being identical. 
Two minima are identical or more precisely `related', if the equilibrium structure of the ionized system lies within the catchment basin of the neutral system and vice versa.Applying this criterion on an extensive database of accurately relaxed geometries, we
arrive at the opposite conclusion.

\section{Methodology}
The global and local minima presented here are obtained within Density functional theory using the`Big DFT' 
wavelet code ~\cite {bigdft} which was coupled to the `minima hopping' ~\cite{minhop} global optimization algorithm.
The local spin density approximation (LDA) is used together with HGH type pseudo potentials~\cite{HGH} for the calculation of the 
potential energy surface. The size of the wavelet basis set was chosen such that the energies were converged to within better than $10^{-4} $ Hartree 
with respect to the infinite size basis set. 
A combination of conjugate gradient and BFGS methods~\cite{bfgs} was used for the local geometry optimizations 
and they were stopped when the numerical noise in the forces 
was about 20 percent of the total force.  This happened usually when the largest force acting on any atom was less than $2\times 10^{-5}$ Hartree/Bohr. 
Saddle points were found by a modified version of the `A spline for your saddle' method~\cite{spline}. 
 
In contrast to plane wave basis sets, free boundary conditions for charged systems are not problematic with a wavelet basis set. 
In plane wave program a neutralizing background charge is needed, since a periodic system can not have a charged unit cell.
In a wavelet basis set the integral equation for the potential $V$ 
$$ V({\bf r}) = \int \frac{ \rho({\bf r}')}{ |{\bf r} - {\bf r}'|} d{\bf r}' $$
can be solved directly for the electronic charge density $\rho$ with a monopole 
and the electrostatic potential can therefore be calculated very accurately for charged systems~\cite{poisson}.

For all the clusters  we have carried out separate global optimization runs for neutral and ionized system.
Since anions with weakly bound additional electrons are less accurately described by density functional theory than 
cations, we considered only cations in addition to the neutral system.
For small clusters ( less than 10 atoms for silicon and less than 20 atoms for magnesium) the majority of low energy 
local minima can be obtained. That this condition is fulfilled can 
be deduced in the minima hopping algorithm from a strong increase in the kinetic energy of the molecular dynamics 
trajectories. For larger clusters this explosion of the kinetic energy~\cite{myreview} can not be observed for 
any reasonable short simulation time. In case of medium sized clusters we calculated always at least 100 low energy local minima structures 
and we did various empirical checks to convince ourselves that the global minimum was found. 
We checked for instance always that the lowest energy structures found for the cation system did not relax upon addition of 
an electron into a structure that was lower in energy than the putative global minimum found for the neutral system. 

Using this approach we investigate whether the global minimum structures of neutral and positively charged clusters are related.
We will use the following two criteria as the definition for two structures of a neutral and ionized system to be ``related''
\begin{itemize}
\item The equilibrium structure $i$ of the cation will relax into the equilibrium structure $j$ of the neutral cluster 
when an electron is added. 
\item The equilibrium structure $j$ of the neutral cluster will relax into the equilibrium structure $i$ of the cation when an electron is removed. 
\end{itemize}
By relaxations we mean local geometry optimization with a sufficiently small step size, which will make it very unlikely that 
the local geometry optimization jumps out of the catchment basin within which the local geometry optimization was started.
The structures of the neutral and ionized system are thus considered  to be related, if there is a one-to-one mapping 
between the global minima structures upon addition and removal of an electron. 
This definition of two structures being related is motivated by the fact that the removal or addition of an electron 
in an experiment is quasi instantaneous on the time scale of the motion of the heavy nuclei. 
A cluster will therefore relax experimentally into the minimum of the catchment basin in which it finds itself 
after the addition or removal of an electron. 
In order to see whether our definition is fulfilled or not, we have introduced mapping charts that show which local 
minimum of the neutral system relaxes into which local minimum of the ionized system and vice versa. 
We consider the global minima structures of the neutral and ionized cluster to be identical if the two global minima structures are 
related according to the above definition. 

In order to detect the degree of similarity between two structures with $N_{at}$ atoms and  atomic coordinates  $R^a$ and $R^b$ respectively 
we have also calculated the configurational distance $D$
$$  D = \frac{1}{N_{at}}  \sqrt{\sum_{i=1}^{3 N_{at}} ({\bf R}^a_i - {\bf R}^b_i) ^2 } $$
The two structure were rotated and shifted in such a way as to minimize $D$. In addition atomic numbers were permutated in the search for the 
smallest possible $D$. 
It turns out that structures, that are related 
according to our definition, usually have also a small configurational distance, but the opposite is not true.

We have chosen silicon and magnesium clusters for this study since they are among the most extensively studied clusters 
and since we wanted to see whether clusters made out of insulating and metallic materials behave in the same way.  

The figures are produced using `v\_sim'(http://inac.cea.fr/L\_Sim/V\_Sim/index.en.html). The symmetry group was found using vmd ~\cite{vmd} plug-ins ~\cite{STON2001}.

\section{Results}
\subsection{Silicon Clusters}
For silicon system we did our calculation for small clusters containing 3 -19 atoms and for $Si_{32}$ as an representative of
medium size clusters. 
For very small clusters there exist only a few local minima structures and they are therefore usually well separated in energy.
As the number of atoms in the cluster grows, the number of meta-stable structures increases exponentially. 
The concept of a global minimum is already rather ill-defined for silicon clusters containing more than some 7 atoms. 
They have many quite distinct structures that are very close in energy to the global minimum structure~\cite{hellmann2}.
As a consequence more than one 
structure can be populated even at room temperature. A second consequence of this is that different density functionals 
can give a different energetic ordering of the various minima~\cite{zheng} and even with the most accurate Quantum Monte Carlo calculations 
it is difficult to obtain the resolution necessary to predict the correct energetic ordering~\cite{hellmann2}. 
In this study we are not claiming to identify the correct ground state structures of the studied silicon clusters, but instead we want to
show general trends. Therefore we use standard density functional theory instead of the extremely expensive Quantum Monte Carlo method. 
Considering the fact that completely different structures can be extremely close in energy suggests strongly that a major perturbation 
such as the addition or removal of an electron can change the energetic ordering of the structures. Older studies have in contrast 
frequently just assumed that the ground state structures of neutral and positively charged clusters are the same. 
\begin{figure}[!h]
\includegraphics[width=0.45\textwidth]{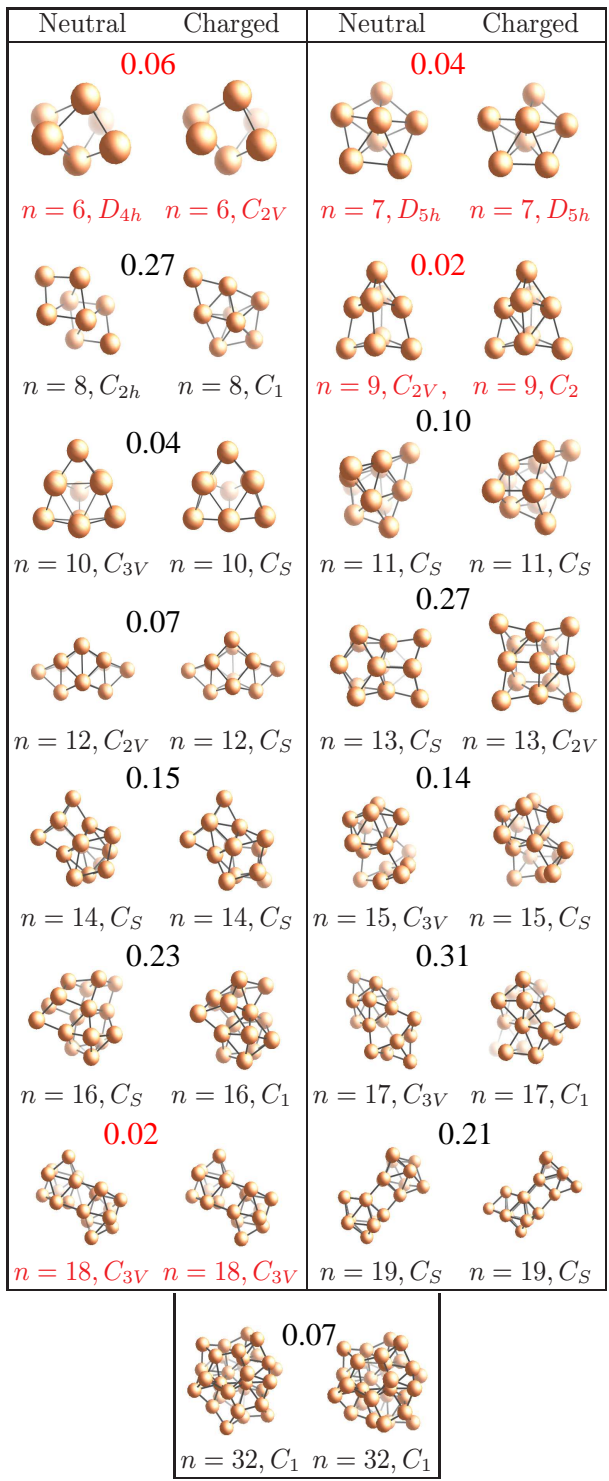}
 \caption{\label{fig:Si_fig} Global minima of charged and neutral $Si_n$, for n=6,7,..19 and 32. Only for n=6,7,9  and 18 the global minima of charged and neutral are ``related''.The configurational distance between each pair is given in \AA.}
 \end{figure}
In some more recent investigations, few cases were identified where the  neutral and positively charged cluster were not `related'. 
In an investigation, where silicon clusters with less than 20 atoms were investigated~\cite{ionization}, $Si_{8}$, $Si_{12}$,$Si_{13}$, $Si_{15}$ and 
$Si_{17}$ were found as the exceptions were the ground state geometries of the cation differ from the one of the neutrals.
In another investigation of silicon clusters with less than 10 atoms~\cite{ionization10}, 
the ground state geometry of $Si_{9}$ and $Si_{10}$ were found to be the ``related''.
Both studies are in contradiction to our results which show that for silicon clusters with more than 7 atoms, the  ground state 
structures of the neutrals and cations are not related with the only exception of $Si_{9}$ and $Si_{18}$ and are as a matter of fact 
quite different( Fig.~\ref{fig:Si_fig}).
In another study of medium sized clusters~\cite{ionization32} it was also found that in most cases the structures of the 
neutrals and cations are the same. Out of the medium size clusters we have only examined the 32 atom cluster for which we however also 
find different ground state structures.
\begin{figure}

\includegraphics[width=0.45\textwidth] {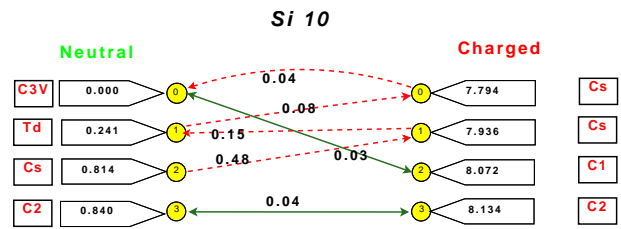}
\caption{\label{fig:Si_10} Mapping chart for $Si_{10}$. The configurational distance between the 
the neutral and charged ground state configurations is very small (0.04 \AA) and  
ionized ground state does relax into the neutral ground state when an electron is added. 
However the neutral ground state does not relax into the ionized ground state
and therefore the structures are not `related' according to our definition. This behavior is rather exceptional and was only found 
for $Si_{10}$ ,$Si_{12}$ ,$Mg_{25}$ and  $Mg_{56}$.  For all the other unrelated structures neither the ionized ground state relaxes 
into the neutral ground state nor the neutral into the ionized one.}
\end{figure}
\begin{figure}
\includegraphics[width=0.45\textwidth]{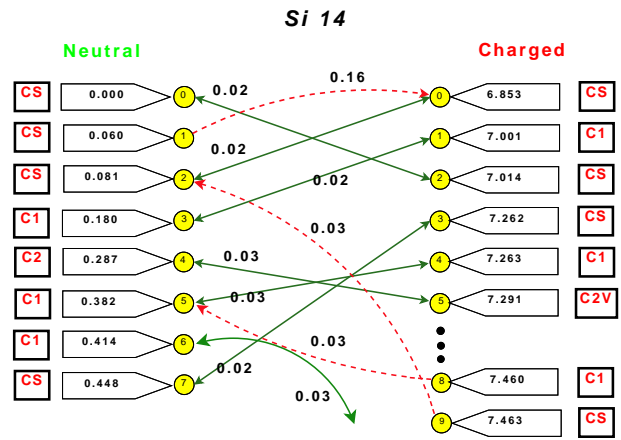}
\caption{\label{fig:Si_14} Mapping chart for $Si_{14}$. The ground state structures are not related and are quite different(FIg:~\ref{fig:Si_fig}).}
\end{figure}

Fig:~\ref{fig:Si_10} and  Fig:~\ref{fig:Si_14} shows the mapping chart which gives detailed information about the relaxation properties upon  addition and removal of an electron .
We distinguish between reversible and irreversible  mappings between pairs of local minima. The energies of all the structures are measured with 
respect to the ground state energy of the neutral system. Solid double arrow connecting lines denote reversible mappings and dashed single 
arrow connecting line irreversible  mappings. The space group is given in the rectangular boxes and the numbers close to the the connecting 
lines give the configurational distance of the two configurations. 
A reversible mapping connects two structures which are related according to our definition. 
In an irreversible mapping, the cluster relaxes from the i-th to the  j-th local  minimum when an electron 
is removed or added, but it relaxes to a structure which is 
different form the i-th when the electron is given back or taken away again. Fig:~\ref{fig:Si_10} and  Fig:~\ref{fig:Si_14} shows that both kinds of mappings are encountered frequently. 
The minima of the neutral and cation are related according to our aforementioned  definition only if a reversible mapping connects the 
two global minima. This case was never encountered for clusters of more than 7 atoms except for $Si_9$ and $Si_{18}$ and the global minimum structures for the neutrals and 
cations are thus different except for  $Si_{n}$ n=3 to 7 , 9 and 18 in this size range . The numerical values along the relaxation 
arrows in the mapping diagrams indicate the configurational distances in the relaxation processes. These distances are typically 
of the order of 0.03 \AA , and thus show that the distortion during the relaxation is rather small. The symmetry group is also 
conserved in most cases. The fact that the geometries change so little upon  removal or addition of an electron might have 
contributed to the wrong believe that the ground state of the neutral and cation are more or less identical. 
Nevertheless these small displacements are frequently sufficient to bring the system in another catchment basin. 

The energetic ordering for neutral and ionized cluster configurations would be identical if the ionization energy or electron affinities 
(including the energy that comes from the small relaxation upon removal or addition of an electron) would be constant, 
i.e. independent of the shape of the various meta-stable configurations. The 
essential point is however that ionization energies and electron affinities are about two orders of magnitude larger than 
the energy differences between the ground state structure and the next meta-stable low energy structures. Relatively small differences in the 
ionization energies and electron affinities between the different configurations can therefore lead to a reversal of the energetic ordering 
of the local minima. The energy differences between the ground state 
and the first meta-stable configuration is of the order of few $k_B T$ at room temperature and the energy differences between 
the higher meta-stable configurations are even smaller.

We find small configurational distance values not only for the structural changes induced by the addition or removal of an electron but also between 
different local minima of the neutral and ionized clusters. 
The configurational distance between the first and second meta-stable configuration of the $Si_{14}$ cluster 
is for instance only 0.15 \AA. Nevertheless the two local minima are separated by a barriers of about 1.2 eV. 
In these disordered structure a broad distribution of barrier heights is to be expected~\cite{lowbar} and we find indeed also 
low barriers.  The configurational distance between the ground state of the charged $Si_{10}$ cluster 
and its first meta-stable configuration is for instance also 0.15 \AA. But the barrier between the two structures are 
much smaller namely 0.22 eV  and 0.08 eV respectively. Such small barrier heights are well below the accuracy level of density 
functional methods and it can hence not be excluded that higher level calculations such as coupled cluster or Quantum Monte 
Carlo calculation would give a different potential energy surface. Our previous experience~\cite{Silandscape} shows however that 
barrier height are quite well reproduced by density functional theory if no bonds are broken during the transformation 
from one structure to the other.

\subsection{Magnesium clusters}    
 
\begin{figure}
\includegraphics[width=0.49\textwidth]{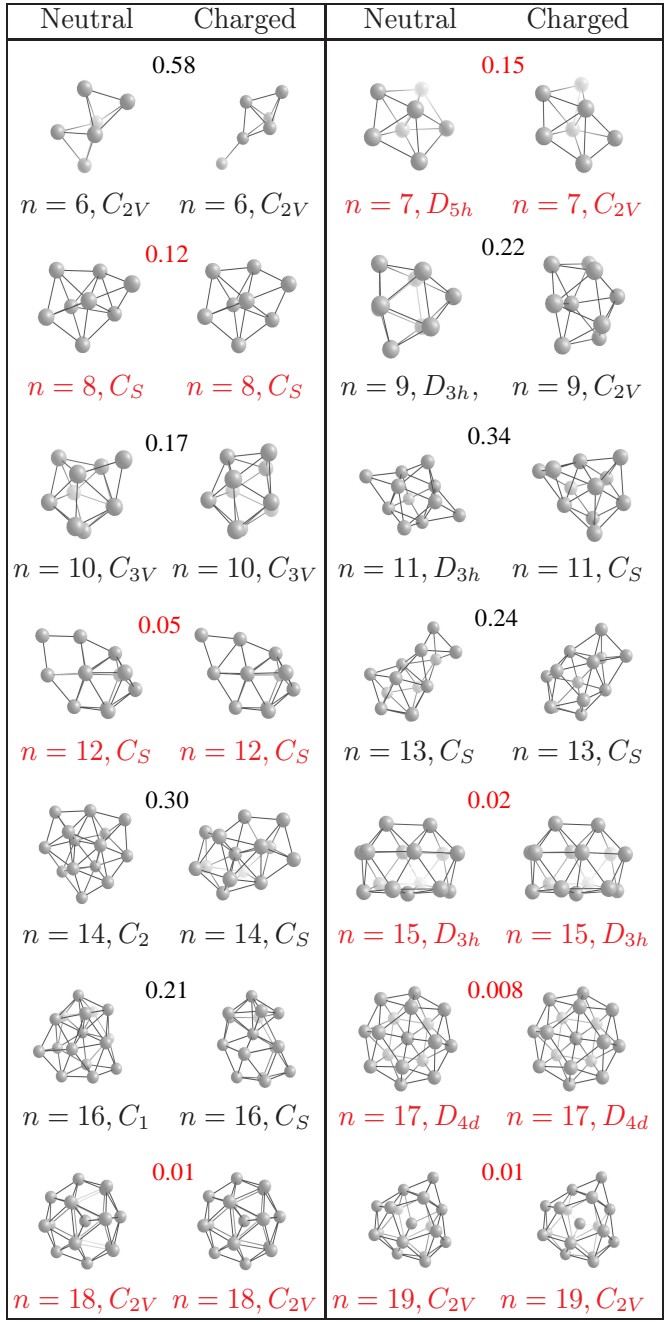} 
 \caption{\label{fig:Mg_fig1} Global minima of charged and neutral $Mg_n$, for n=6-19. Only for n=7,8,12,15,17,18 and 19 the global minima of charged and neutral are ``related''.
The configurational distance between each pair is given in \AA. }
 \end{figure}

\begin{figure}
\includegraphics[width=0.48\textwidth]{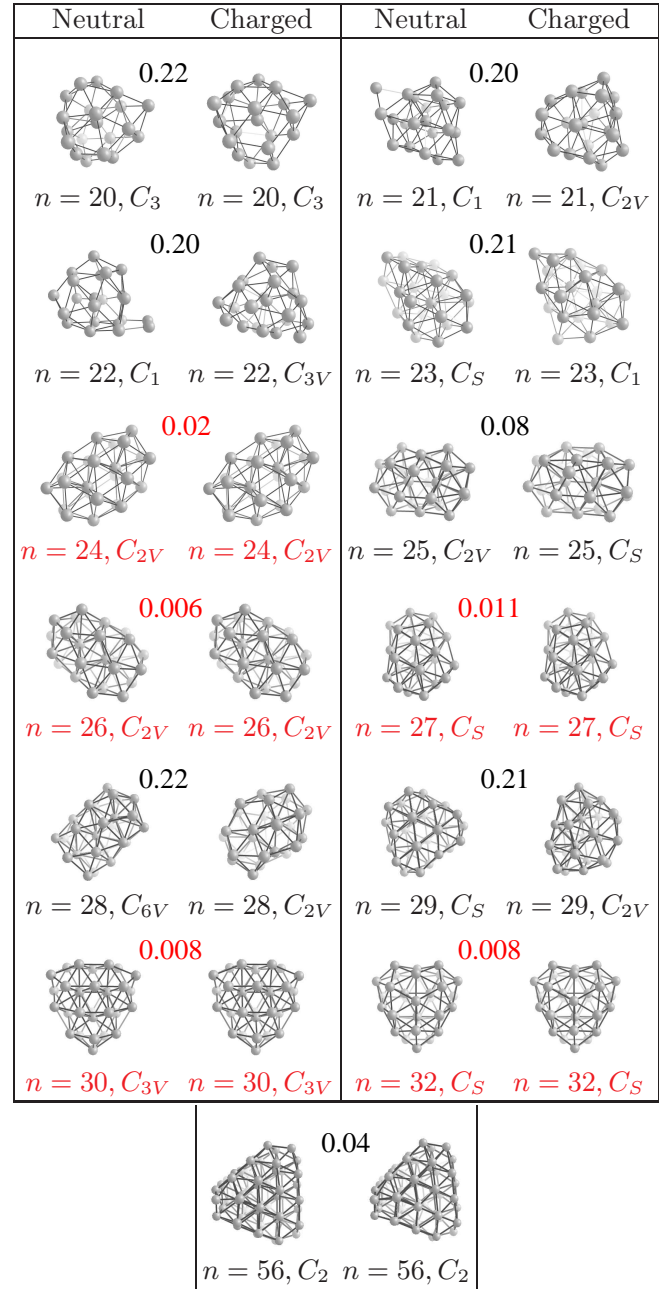}
\caption{\label{fig:Mg_fig2} Global minima of charged and neutral $Mg_n$, for n=20-30,32,56. Only for n=24,26,27,30 and 32 the global minima of charged and neutral are ``related''.The configurational distance between each pair is given in \AA }
\end{figure}

For $Mg_{n}$  we have systematically studied all small and medium size clusters with n=6 to 30 atoms as well as $Mg_{56}$.The Global minima are shown in Fig.~\ref{fig:Mg_fig1} and
 Fig.~\ref{fig:Mg_fig2}.
For these cluster sizes 
the electronic HOMO-LUMO gap does not yet tend to zero, but is around 0.1 eV. 
So no pronounced metallic behavior is present. The ionization energies are also comparable to the case of the silicon clusters. 
The ionization energy  is on average 5 eV for the magnesium clusters and 7eV for the silicon clusters. 
The only notable difference we found between the silicon and magnesium cluster is number of meta-stable states,
which is much larger for silicon clusters. Since all energy differences are however also smaller for $Mg$ than for $Si$  ,the 
average configurational distance between different meta-stable configurations is again similar in both cases.
Hence $Mg$ clusters have the same overall behavior as the $Si$ clusters, i.e in general the neutral and ionized ground 
states are not related.

\begin{figure}
\includegraphics[width=0.49\textwidth]{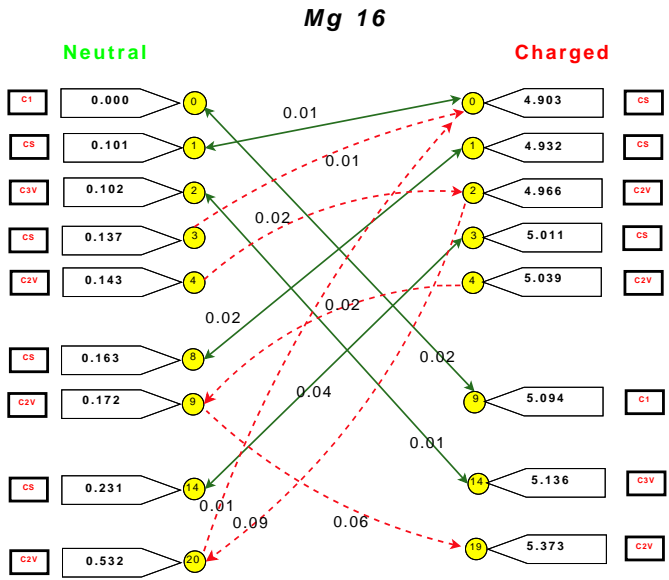}
\caption{\label{fig:Mg_16} Mapping chart for $Mg_{16}$. The ground state of the neutral cluster is mapped to a rather high local minimum of the charged 
cluster. }
\end{figure}

\begin{figure}
\includegraphics[width=0.49\textwidth]{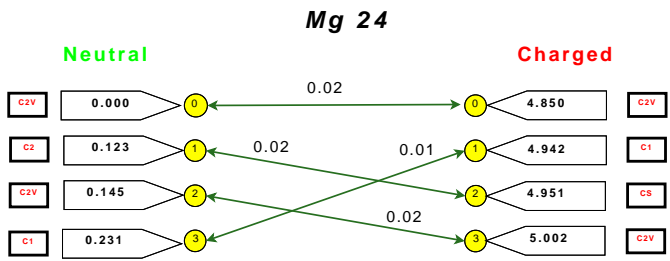}
\caption{\label{fig:Mg_24} Mapping chart for $Mg_{24}$. For this system the ground states are related. The higher energy meta-stable 
states are however even for such a system typically not `related'. }
\end{figure}

In the studied size range we find that the global minima of neutral and cation clusters are related for n=7,8,12,15,17,18,19,24,26,27,30 and 32. 
For a bigger system ,$Mg_{56}$ we also found the global minima to be different for charged and neutral system.
So in total the ground state structures are related in 12 cases and unrelated in 21 cases.
Fig.~\ref{fig:Mg_16} and Fig.~\ref{fig:Mg_24} exemplifies the same kind of mapping for $Mg_{16}$ and $Mg_{24}$   between charged and neutral system as we already 
showed for silicon systems. These mapping charts 
look very similar to that of silicon systems, i.e. the energetic ordering changes when the system goes from the neutral to the charged state. Although for $Mg_{24}$ the neutral and
charged global minima are `related' , from the mapping chart (Fig.~\ref{fig:Mg_24}) we can see the  sign of energetic ordering changes in the system while going from neutral to charged
state. 
The numerical values along the relaxation 
arrows in the mapping diagrams indicate the configurational distances  in the relaxation processes. These distances are typically 
of the order of 0.02 \AA  , unlike Silicon systems where this value is 0.03  \AA ,  and thus show that the distortion during the relaxation is smaller than that of silicon systems.

Our results are again overall in disagreement with the majority of previous publications. 
In one of the earliest publication on this topic, where 
clusters with up to 6 atoms were studied, identical ground state structures were found for $Mg_{6}$ and $Mg_{7}$~\cite{reuse}. 
In a study of Mg cluster with up to 21 atoms, it was found 
that only for $Mg_{3}$ and $Mg_{4}$ the ground states are different~\cite{lyalin}.  In another somewhat more extensive study in the range 
between 2 and 22 atoms~\cite{jellinek}, it was found that in addition also 
$Mg_{6}$, $Mg_{7}$, $Mg_{8}$, $Mg_{11}$, $Mg_{12}$ and $Mg_{13}$ have different ground states. 

We have also recalculated the energetic ordering of the minima of several magnesium clusters with the PBE functional~\cite{PBE}. In all these cases the 
ordering was identical to the ordering with the LSD functional. This is in contrast to the silicon clusters where the energetic ordering depends on 
the functional being used. This suggests that the density functional results for the magnesium clusters are very reliable.

For the magnesium clusters the average configurational distance between the various local minima is typically in the range between 0.1 \AA , and 
0.2 \AA , and thus larger than the average configurational distance of the relaxation induced the the removal or addition of electrons. 
Since the magnesium cluster are also disordered we find, as in the case of the silicon clusters, a broad distribution of barrier heights.
We calculated randomly 12 barrier heights of the neutral $Mg_{16}$ and $Mg_{24}$ cluster and we found values in between 0.05 and 0.8 eV. 


 \section{Conclusion}    
Using an exhaustive sampling of the low energy configurations based on the minima hopping method we show for silicon and magnesium clusters that 
the ground states of neutral and ionized clusters are in general not related and are in many cases quite different. 
This comes from the fact that for medium and large clusters there are in general 
numerous meta-stable structures which are energetically very close to the ground state. The differences in ionization energies and 
electron affinities for  different structures are much larger than this energy difference between structures.
These facts have to be taken into account in the interpretation of experiments with ionized clusters. 

There is no reason to believe that clusters made out of other elements behave differently. Based on our arguments 
one can only expect that for certain magic cluster sizes, for which ground state structures exist that are considerably lower in 
energy than other competing meta-stable structures, the ground state does not change upon removal or addition of an electron. 
Such an example is for instance the $C_{60}$ fullerene. 
\section{Acknowledgment}
We thank the Indo-Swiss Research grant and SNF for the financial support  and the
CSCS for computer facility. 
  

\end{document}